%% file: HybridSmoothingSurfaces.tex
\documentclass[letterpaper, 10 pt, conference]{ieeeconf}  
\usepackage{amsmath}
\usepackage{amssymb}
\usepackage{graphicx}
\usepackage{enumerate}
\usepackage{multirow}
\usepackage{url} 
\usepackage[table]{xcolor}
\usepackage{verbatim}
\usepackage{float} 
\usepackage{setspace}
\usepackage{tablefootnote}
\usepackage{wrapfig}
\include{ourDefinitionsLite.tex}

\IEEEoverridecommandlockouts                              
\title{\LARGE \bf
Hybrid Bayesian Smoothing on Surfaces
}

\author{Matthew Hofkes\hspace{.2cm}\\
     Department of Statistics, Colorado School of Mines\\
     and \\
     Douglas Nychka \\
     Department of Statistics, Colorado School of Mines}

\begin{document}

\maketitle
\thispagestyle{empty}
\pagestyle{empty}

\begin{abstract}

Modeling spatial processes that exhibit both smooth and rough features poses a significant challenge.  This is especially true in fields where complex physical variables are observed across spatial domains. Traditional spatial techniques, such as Gaussian processes (GPs), are ill-suited to capture sharp transitions and discontinuities in spatial fields. In this paper, we propose a new approach incorporating non-Gaussian processes (NGPs) into a hybrid model which identifies both smooth and rough components. Specifically, we model the rough process using scaled mixtures of Gaussian distributions in a Bayesian hierarchical model (BHM). 

Our motivation comes from the Community Earth System Model Large Ensemble (CESM-LE), where we seek to emulate climate sensitivity fields that exhibit complex spatial patterns, including abrupt transitions at ocean-land boundaries. We demonstrate that traditional GP models fail to capture such abrupt changes and that our proposed hybrid model, implemented through a full Gibbs sampler. This significantly improves model interpretability and accurate recovery of process parameters.

Through a multi-factor simulation study, we evaluate the performance of several scaled mixtures designed to model the rough process. The results highlight the advantages of using these heavier tailed priors as a replacement to the Bayesian fused LASSO. One prior in particular, the normal Jeffrey's prior stands above the rest.  We apply our model to the CESM-LE dataset, demonstrating its ability to better represent the mean function and its uncertainty in climate sensitivity fields.

This work combines the strengths of GPs for smooth processes with the flexibility of NGPs for abrupt changes. We provide a computationally efficient Gibbs sampler and include additional strategies for accelerating Monte Carlo Markov Chain (MCMC) sampling. 

\end{abstract}

\section{INTRODUCTION}

In many areas of data science, observations collected over a continuous spatial domain represent a surface.  The goal of spatial statistics is to model this function and predict unobserved locations. In the environmental and geosciences, it is common to encounter data where physical variables are observed or simulated across a spatial domain, with the statistical properties of these fields being of scientific interest.  Spatial statistics offers a rich set of models to address such problems.  The models are extensively covered in Noel Cressie’s Statistics for Spatial Data \cite{cressie2015statistics} and further developed through Bayesian hierarchical models \cite{wikle2010general, kang2011bayesian, gelfand2017bayesian}, stochastic partial differential equations (SPDE) \cite{holden1996stochastic}, and computational strategies for large data volumes \cite{katzfuss2017multi, katzfuss2012bayesian, zammit2017frk}.  More recent work addressing spatial extremes includes \cite{huser2022advances}.  The focus of these efforts, while being innovative and resourceful,  has been on Gaussian process (GP) models to represent processes that exhibit some continuity across space. In this work, we propose an addition to this modeling framework which involves two spatial processes simultaneously modeling abrupt and smooth behavior in the spatial field.  This is accomplished by considering the spatial field which is the sum of  Gaussian and non-Gaussian processes, with the non-Gaussian component built in stages using a Bayesian hierarchical model (BHM). The non-Gaussian component takes inspiration from the Bayesian fused LASSO   which is represented as a scaled mixture of Gaussian variables \cite{casella2010penalized}. However, the limits of this model are well know, and we extend this to a class of other scaled mixtures. 

This work is motivated by the challenges of statistical modeling related to the Community Earth System Model Large Ensemble (CESM-LE) \cite{kay2015community}. These simulations currently consist of approximately 30 ensemble members that capture the Earth's internal (natural) climate variability.  A key data science challenge is to efficiently emulate subsets of the model's variables, enhancing the sample of input fields for climate impact assessments.   As an example, we focus on the climate sensitivity of a location to global temperature changes. A prominent feature in this geophysical field is the stark ocean/land difference, separate from smoother spatial dependencies. This feature motivates the need for representing both rough and smooth components. A traditional spatial statistics approach might employ a GP for climate sensitivity and include an ocean/land indicator with a regression model to account for this boundary. However, such a model fails in cases where the boundary is not well defined. Even in this example, the transition between ocean and land is subtle and not always aligned with coastlines, making an ocean/land covariate inadequate.  To address this, we developed a non-Gaussian process (NGP) model that allows for areas where the field is constant but also builds in abrupt jumps similar to what we observe at the ocean/land boundaries. This addition enables us to simulate climate sensitivity fields that are more realistic of the large ensemble (LENS) simulations. 

Traditional Kriging and non-parametric spatial models are ill-equipped to handle data that are a combination of smooth processes and discontinuities such as those observed in the LENS sensitivity field and common in other geophysical fields.  (For brevity we refer to this second component as the {\it rough} process although we  acknowledge that there can many different ways  non-Gaussian process can exhibit ``roughness".)   If covariates are known, universal Kriging has the potential to identify step changes as a regression component; however, in many cases this covariate information is not available. Here we take the alternative approach by defining a separate spatial process.  In this work, we define a NGP in terms of operations that convert the field into scale mixtures of  independent normal random variables.  This perspective has roots in the seminal work by Lindgren \cite{lindgren2011explicit} where the Gaussian spatial model is defined from a process, SPDE point of view and the resulting covariance function being a secondary aspect. In addition, we draw on the more recent work of expanding fields in basis functions \cite{hofkes2024hybrid}, \cite{nychka2019package}, and \cite{wiens2020modeling} as a  way to discretize this model for practical implementation.  An important feature of our model is to represent the rough process as a scale mixture of GPs. In this way, we can leverage the advantages of Bayesian computation for multivariate normal distributions, while also describing processes that are very different from GP realizations.

In our previous work considering anomaly detection in time series  \cite{hofkes2024hybrid}, we used a cubic smoothing spline for the smooth signal and the Laplace distribution prior to describe non-Gaussian step changes (i.e. change points) in the time series. We refer to this as ``hybrid smoothing", and we prefer a Bayesian implementation based on a BHM. To extend hybrid smoothing to a spatial context, we have made two changes. First, we focus on a model for the rough process directly, as sketched above, instead of expanding the process with basis functions. This reduces the complexity of developing methods and is reasonable given that the climate output is on a regular grid. 

Second, we evaluate several different choices for priors on $\blambda^2$ as alternatives to the exponential distribution associated with the Laplace. We demonstrate that these other priors, with heavier tails, are successful where the Bayesian fused LASSO fails.  They produce better realizations of what we expect for ocean/land boundaries and also better posterior interpretations. 
 
The implementation of a full Gibbs sampler for the BHM is a key feature of this paper.  We employ a number of techniques including orthogonalization, adaptive burn-in, and partial updates to speed computation.  The net result is a strategy that minimizes the number of MCMC iterations to fit the hybrid model.  To showcase this implementation, for the rough spatial process, we explore five scaled mixtures and detail the posteriors associated with each.

In Section \ref{BHM}, we discuss the BHM, as well as the construction of the NGP. Section \ref{Priors} expands upon Section \ref{BHM}, taking a deeper look at five Gaussian mixtures used to construct the NGP and exploring their effects relative to a normal prior.  Additionally, this section tests the ability of these NGPs to simulate fields with rough step features.

In Section \ref{simstudy}, we discuss the results of a multi-factor simulation study comparing the performances of five scaling priors with a thin plate spline (TPS) on levels comparable to what is observed in the CESM-LE.  

In Section \ref{Implementation}, we discuss the implementation of the model.  Details are included regarding orthogonalization of the fixed and spatial effects, using an adaptive approach to prevent rapid fusion in the rough function, and speeding up the MCMC with partial updates from the posterior. Section \ref{Gibbs} includes the full Gibbs Sampler for the model and its possible posteriors depending on the scaled mixture. 

Section \ref{CSD} details the application of the model to the CESM-LE, presenting an alternative estimate of the mean function and uncertainty. We illustrate the significant improvement in uncertainty and effective degrees of freedom when using the hybrid model. 

Finally, Sections \ref{Extensions} and \ref{Conclusion} include extensions to the model, additional research directions, and a summary of our work.


\section{Bayesian Hybrid Smoothing of Surfaces}\label{BHM}

We propose the following hierarchical model as a simple and natural way to extend Kriging on a grid to surfaces with abrupt step changes.     We assume that all variables are observed at  the locations on a regular  two-dimensional grid and  consider the full  additive model for the observations as 
\[ \bz = X\bbeta + \by + \bgamma +  \bepsilon \]
where $\bbeta$ are regression parameters from a linear model,  $\by$ is a smooth Gaussian  component, $\bgamma$ a potential discontinuous or rough surface, and $\bepsilon$ independent normal errors or Gaussian white noise. Starting with this as the observation layer,  we add the following levels to fill out a BHM.  

\begin{align}\label{int.BHM}
    \bz|\bbeta,\by,\bgamma,\tau^2 &\sim N(X\bbeta+\by+\bgamma,I\tau^2)  \\
    \by|\sigma^2&\sim N(0,K\sigma^2) \text{ where $K$ is a GP covariance.}\nonumber \\
    \bgamma|F(\blambda^2) &\sim N(0,F(\blambda^2)) \nonumber \\
    \pi(\blambda^2) &  \text{ as a scaling distribution}\nonumber\\
    [\bbeta] &\propto 1 \nonumber\\
    \tau^2 & \sim \text{inv-gamma}(\alpha_{\tau^2},\beta_{\tau^2})  \nonumber \\
    \sigma^2 & \sim \text{inv-gamma}(\alpha_{\sigma^2},\beta_{\sigma^2})  \nonumber
\end{align}

\subsection{The Construction of $Q(\blambda^2)$ and $F(\blambda^2)$}
A key contribution of this model is expressing the rough function as a scale mixture of multivariate normals.   The covariance matrix for this mixture is what encodes the spatial structure of  the $\bgamma$ component.  Here we  describe the construction of $F(\blambda^2)$  based on  a two-dimensional grid. We will see this has the form of a Markov random field, similar to the one dimensional case. 

For a regular two dimensional grid let $m$ be the number of unique pairs of grid points that are adjacent,  i.e., all pairs are nearest neighbors. Let $\nu$  and the functions $i(\nu)$ and $j(\nu)$ index these pairs  so the field values $\bgamma_{ i(\nu)}$ and $\bgamma_{ j(\nu)}$ are a nearest neighbor pair.  With this notation, let $g(\bgamma)$ be the pdf for  the joint distribution of $\bgamma$. Expressed as a normal scale mixture it has the  proportionality  
\[ g(\bgamma) \propto \pi^*(\bgamma) \int exp\left( -\frac{1}{2}  \sum_{\nu=1}^m \frac{( \bgamma_{i(\nu)}- \bgamma_{j(\nu)})^2}{\blambda_\nu ^2}\right) \pi(\blambda^2) d\blambda^2   \]
where the prior $\pi^*(\bgamma)$ contains information to make $\bgamma$ identifiable.  This information corresponds to the constraints that make $Q$ a valid precision and is discussed in Section \ref{validQ}.   Additionally, $\pi$ is an $m$ dimensional  scaling distribution for $\blambda^2$. 
Now let $D$ be an $m\times n$  differencing matrix so that $D_{\nu, i(\nu)} =1$ and $D_{\nu, j(\nu)} =-1$ and with the other  entries of $D$ being zero. Simple linear algebra results in 
\[ \sum_{\nu=1}^m \frac{1}{\lambda_\nu ^2}( \bgamma_{i(\nu)}- \bgamma_{j(\nu)})^2  = \bgamma^T D^T \Lambda^{-1} D \bgamma \]
with $\Lambda$ a diagonal matrix with entries $\blambda^2$  and so 
\[ g(\bgamma) \propto  \pi^*(\bgamma) \int exp\left( -\frac{1}{2} \bgamma^T ( D^T \Lambda^{-1} D )\bgamma \right) \pi(\blambda^2) d\blambda^2  \]

We can identify the expression in $\bgamma$ as proportional to a multivariate normal  distribution with precision matrix 
\begin{equation}\label{Qdef}
Q(\blambda^2) = D^T \Lambda^{-1} D + E
\end{equation} 
\noindent where $E$ contains the information conveyed in $\pi^*(\bgamma)$. With these definitions,

\[ [\bgamma] =  \int  [\bgamma | \blambda^2] [\blambda^2] d\blambda^2 \] 
and   
\[ [\bgamma | \blambda^2] = MN(0, F(\blambda^2) ). \]
where $F = Q^{-1}$.
Although not required for our application, we note  that $Q$ is a sparse matrix with at most 5 nonzero elements in each row.  

This approach is similar to the one used by Kakikawa \cite{kakikawa2023bayesian} in applying the Bayesian fused LASSO to linear regression.  For a common value across all $\lambda^2$s, $Q$ is the graph Laplacian.  It is also the precision matrix used in a SAR1 model, and the square root of the precision used in LatticeKrig, a sparse approximation method for large data \cite{nychka2019package}.

\subsection{Choosing $\pi(\blambda^2$)}\label{choosinglambda}

The NGP, $\bgamma$,  is expressed as a scaled mixture of Gaussian distributions. We aim for this mixture to induce sparsity in differences between adjacent $\bgamma$ values, such that these differences are nonzero only at regional boundaries. Andrews \cite{andrews1974scale} demonstrated that many heavy-tailed distributions, which are sparsity-inducing priors, can be represented as scaled mixtures of Gaussians. This hierarchical formulation is advantageous because it converts a single Metropolis Hastings (MH) update into two computationally efficient Gibbs updates.

While the LASSO is a very effective frequentist technique, the Laplace distribution is not as effective in BHMs \cite{park2008bayesian}, \cite{polson2010shrink}, and \cite{piironen2017sparsity}.  The scaled mixture format allows for the use of other sparsity inducing priors.  One notable mixture is the Horseshoe prior \cite{carvalho2010horseshoe} and \cite{piironen2017sparsity}, which was designed for coefficient shrinkage in linear regression and is widely used in BHMs. The horseshoe prior effectively shrinks small coefficients absolutely while leaving larger ones relatively untouched.  The term ``horseshoe" now broadly refers to a range of priors, all including two half-Cauchy distributions.  One commonly used horseshoe prior includes half-Cauchy hyperparameters for both individual and global shrinkage.   

In this work, we use the original horseshoe, which consists of a half-Cauchy scaled mixture of half-Cauchy scaled mixtures of Gaussians instead of the individual/global setup.  We do this for two reasons.  First, unlike regression where coefficient shrinkage is generally positive, no such benefit is associated with global shrinkage in a spatial setting.  Second, the global shrinkage parameter is widely known to be difficult to infer and often unstable. 

Another scaled mixture prior is the Normal Jeffrey's (NJ) prior which was first introduced in \cite{bernardo1979reference}.  Since then, it has been the subject of many works, particularly in image processing in combination with wavelets \cite{figueiredo2001wavelet}, \cite{1227989}, and \cite{4799134}.  The NJ prior has been lauded as ``universal" and ``objective" due to its lack of hyperparameters.  It is notable for being an improper prior and its ability to shrink small coefficients absolutely while applying almost zero shrinkage to larger ones. 

Additionally, we consider two other scaled mixtures: an inverse gamma-scaled mixture resulting in a Cauchy prior on the differences, and a Pareto-scaled mixture. Both employ heavy-tailed scaling distributions to create priors that emphasize both absolute shrinkage and minimal shrinkage around zero for the differences.
	
	\begin{table}
  		\centering
  		\caption{Scaled Gaussian priors for \textcolor{magenta}{$\bgamma_{i(\nu)}-\bgamma_{j(\nu)}$} values.}
  		\renewcommand{\arraystretch}{1.4} 
  		\begin{tabular}{|c|c|c|}
   	 	\hline
		\textbf{Scaling }  & $\pi(\blambda)$  \textbf{Prior} & \textbf{Results in} \\
      		\textbf{ Mixture}  &  & \\ \hline \hline
      		Exponential & $\lambda^2_{\nu}\sim \text{exp}(\frac{b^2}{2})$ 	 & Laplace$(0,\frac{1}{b})$	\\ \hline
      		HalfCauchy/ & $\lambda_{\nu} \sim C^+(0,t)$ & Horseshoe \\ 
		HalfCauchy & $t\sim C^+(0,b)$  & \\ \hline
      		InvGamma& $\lambda^2_{\nu}\sim \text{IG}(\frac{1}{2},\frac{1}{2b^2})$  & Cauchy$(0,b)$  \\  \hline
      		Jeffrey's   & $[\lambda_{\nu}^2] \propto \frac{1}{\lambda_{\nu}^2}$ & Normal Jeffrey's \\ \hline
      		Pareto & $\lambda^2_{\nu}\sim \text{Par}(\alpha, \lambda^2_{min})$ & Normal Pareto \\ \hline
 		 \end{tabular}
		 \label{scaled}
	\end{table}

\section{Comparing the Scaling Priors in the NGP}\label{Priors} 

\subsection{The Thresholding Effect of Each Prior}

\begin{figure}
    \centering
    \includegraphics[scale=.2]{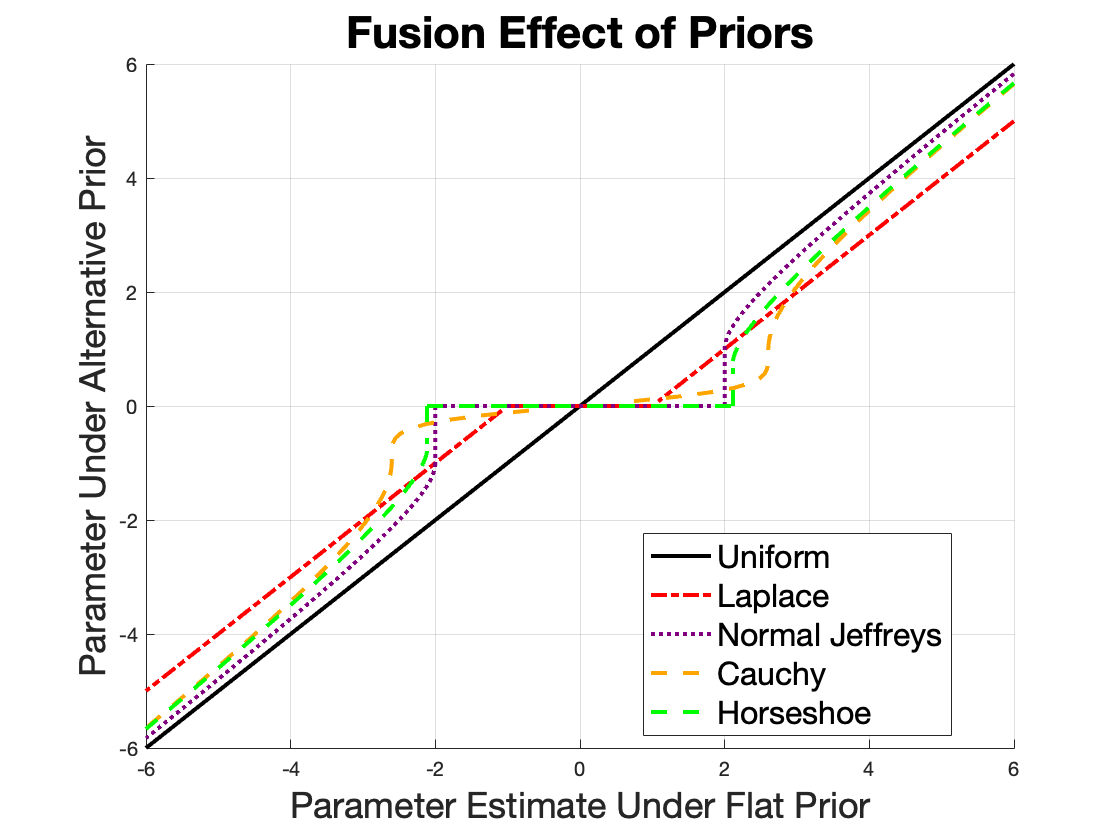}
    \caption{The fusion effects associated with each prior, while setting $\tau^2=1$. The scaling parameters for the Laplace and Cauchy priors are $1$ and $2$ respectively.  The scale on the lowest level of the horseshoe is $1$.   }
        \label{thresholding.effects}
\end{figure}

It is helpful to consider the effects of these priors on the adjacent differences in $\bgamma$ relative to what they would be under a uniform prior.  To assess the effects, we use a method similar to the one described by Polson \cite{polson1991representation}.  This method is effectively the equivalent of the proximal step (also known as the thresholding step) in proximal gradient descent algorithms such as ISTA \cite{daubechies2004iterative} and FISTA \cite{beck2009fast}. This step is expressed over each parameter independently with the optimization problem:
\begin{align}\label{score2}
\begin{array}{rclcl}
\displaystyle \argmin_{\theta} & \multicolumn{3}{l}{\frac{1}{2}(\theta-\theta^*)^2 +P(\theta)}
\end{array}
\end{align}

where $P(x)$ is the penalty added to the cost function, and $\theta^*$ is the update following the gradient descent step. 

This step can be interpreted statistically as computing the shift in a parameter value that occurs when a prior distribution is introduced, compared to the benchmark $\theta \sim N(\theta^*,1)$ where $\theta^*$ is the posterior mean under a uniform prior. By combining the benchmark distribution with the prior distribution and setting the score function to zero, we arrive at the equivalent statistical solution to Equation \ref{score2}.

\begin{align}
	E(\theta | \theta^*) = \theta^* + \frac{d}{d\theta} log\left( \int p(\theta | \alpha ) \pi ( \alpha )d \alpha \right)
\end{align} 
where $\alpha$ is the set of hyper-parameters in the prior distribution of $\theta$. 

Choosing $\theta = \bgamma_{i(\nu)}-\bgamma_{j(\nu)}$, we see the effects associated with each prior illustrated in Figure \ref{thresholding.effects}. The Laplace distribution closely tracks the soft thresholding operator, while the other priors align more closely to a hard thresholding operator. Given that the rough function should ideally achieve absolute shrinkage for small differences and no shrinkage for larger ones, these alternatives to the Bayesian fused LASSO are appealing. The normal Jeffrey's mixture is the most extreme of the four in accomplishing this. 

\subsection{Simulating Non-Gaussian Random Fields}

\begin{wrapfigure}{r}{0.3\textwidth}
  \centering
  \includegraphics[scale=.5]{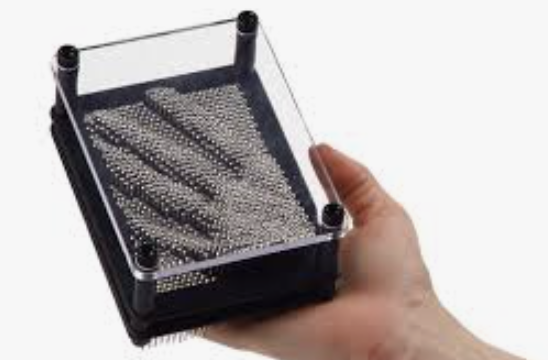}
\end{wrapfigure}

Some spatial process have the ability to generate spatial fields that look like the data they model. This is true of GPs, for example.  Other spatial processes and priors have the ability to fit data but do not the process ability to generate a desired structure without the influence of the data.  Think, for example, of pin impression toys.  The pins can model almost anything, but the pins themselves are not disposed to create hand impressions without the help of a human hand.  In this section, we test whether any of the NGP mixtures are able to produce rough step changes without the influence of data.

 
The rough spatial process is a scaled mixture of Gaussian distributions, so we can simulate these NGP similar to how GPs are simulated.  This is done by first simulating $\blambda^2$ from the scaling distribution and constructing $Q(\blambda^2)$.  Then the square root of $F$ is multiplied by a vector of i.i.d. $N(0,1)$ random variables.  This Gaussian white noise is held steady in all of the simulated fields.

\begin{figure}
    \centering
    \includegraphics[scale=.22]{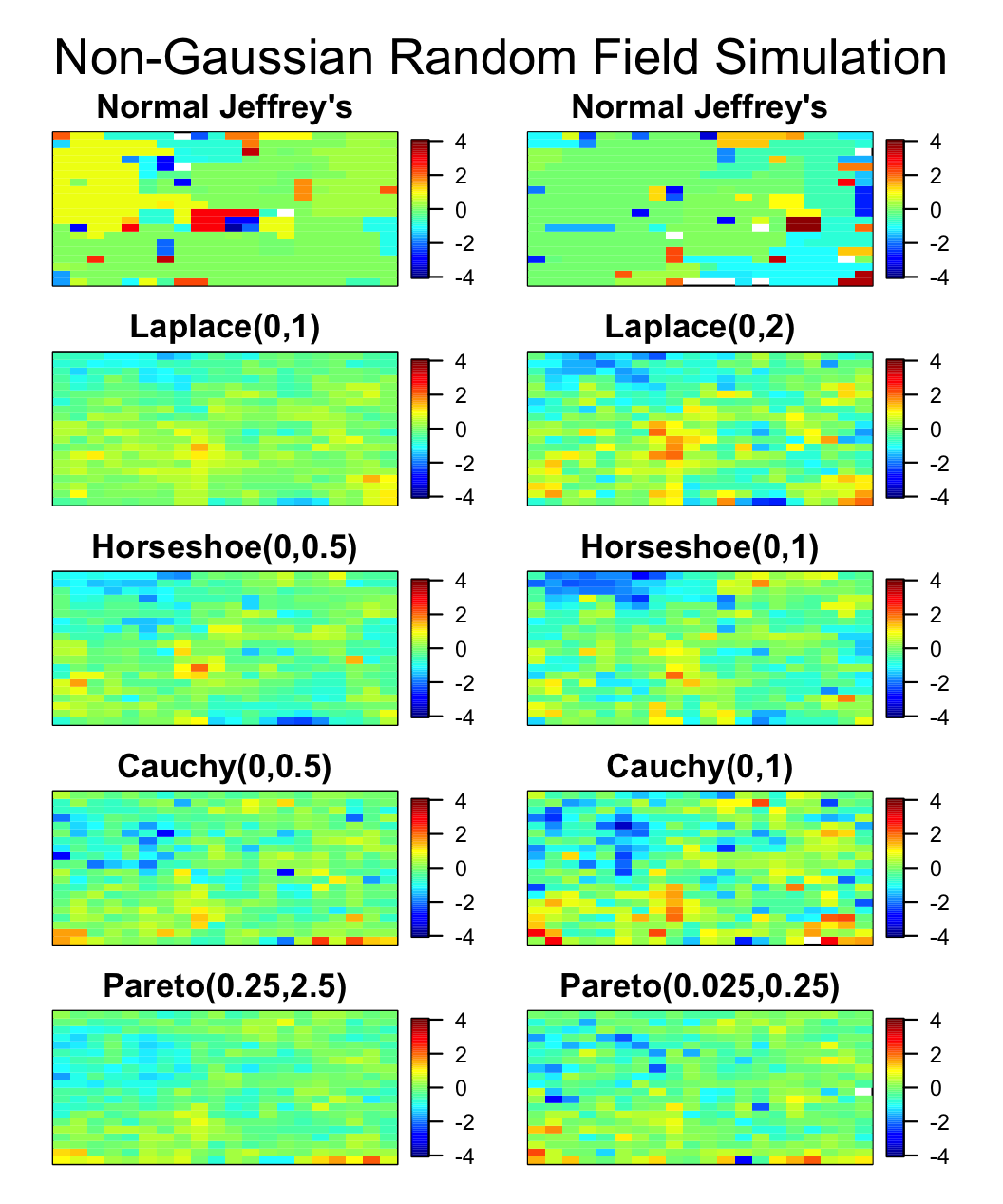}
    \caption{The simulation of Non-Gaussian fields under a variety of mixture priors.  The NJ prior has no hyper parameters, so the two realizations are the same process. The Horseshoe($ \cdot $) refers to the scale parameter in the second layer of the scaling distribution. }
        \label{RandomFieldSimulation}
\end{figure}

The spatial fields from Laplace, Horseshoe, Pareto Scaled Mixture, and Cauchy priors all contain outlier-like features; however, they appear to lack the capacity to produce fields that actually look like step changes in the absence of data. Each of the distributions has hyper parameters, and three combinations are chosen for each process to produce fields that give the reader a general sense of what is possible.  These fields are included in Figure \ref{RandomFieldSimulation}.  The NJ prior, on the other hand, appears to have the innate ability to produce spatial fields that look like step changes.  This is demonstrated in three simulated fields from the NJ also found in Figure \ref{RandomFieldSimulation}.  We discuss sampling from the improper Jeffrey's prior in Appendix \ref{jeffreys}.

It is clear in Figure \ref{RandomFieldSimulation} that the NJ prior stands out in its ability to simulate step changes in the absence of data.

\section{Simulation Study}\label{simstudy}

To test the effectiveness of the hybrid spatial model, we implement a simulation study as a $3\times4\times6$ factorial design using three levels of noise: $\tau^2 $= $0.001$, $0.01$, and $0.1$, four magnitudes of the rough function: $.5$, $1$, $2$, and $4$ units, and six models: the five sparsity inducing priors found in Table \ref{scaled} and a Bayesian implementation the TPS using an MCMC to identify the GP parameters.  We choose these levels to correspond to be relevant to the data fields in CESM-LE.  The rough step changes were generated to mimic coastlines and establish a basis for using this method on the CESM-LE.  For each combination of noise and magnitude levels, $100$ realizations of the spatial data are fit using the six models.

\begin{figure}
    \includegraphics[scale=.13]{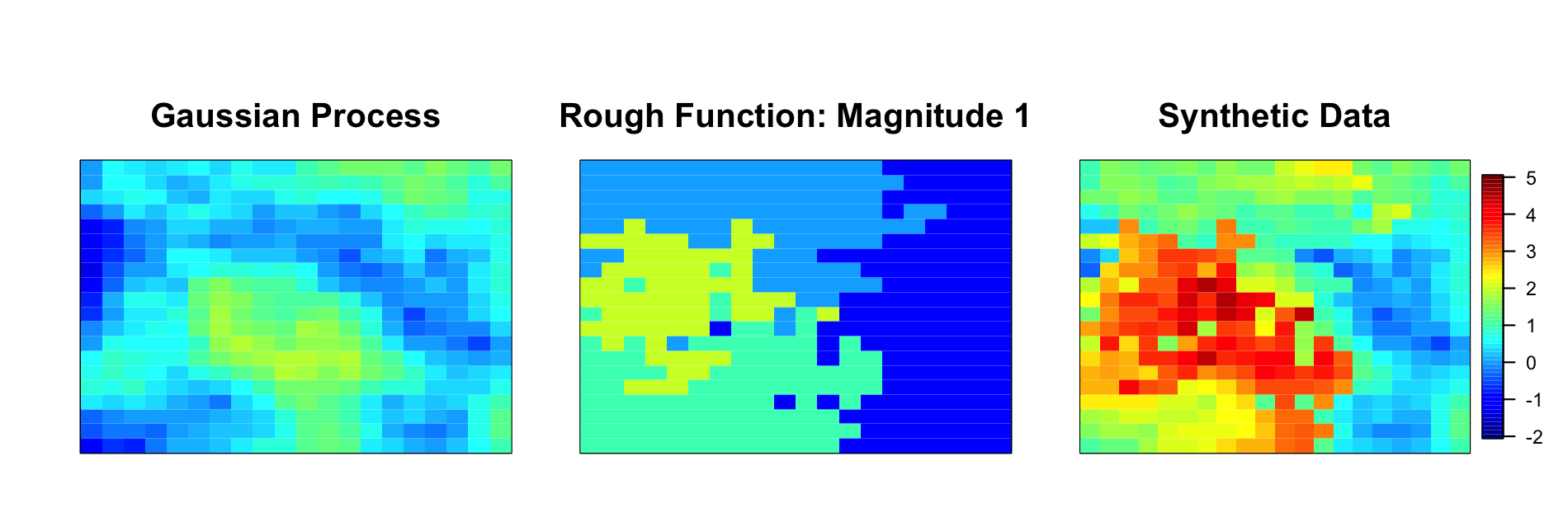}
    \caption{An example of combining a GP and rough NGP to build synthetic data for the simulation study.}
    \label{synthetic.data}
\end{figure}

\begin{figure}
    \centering
    \includegraphics[scale=.16]{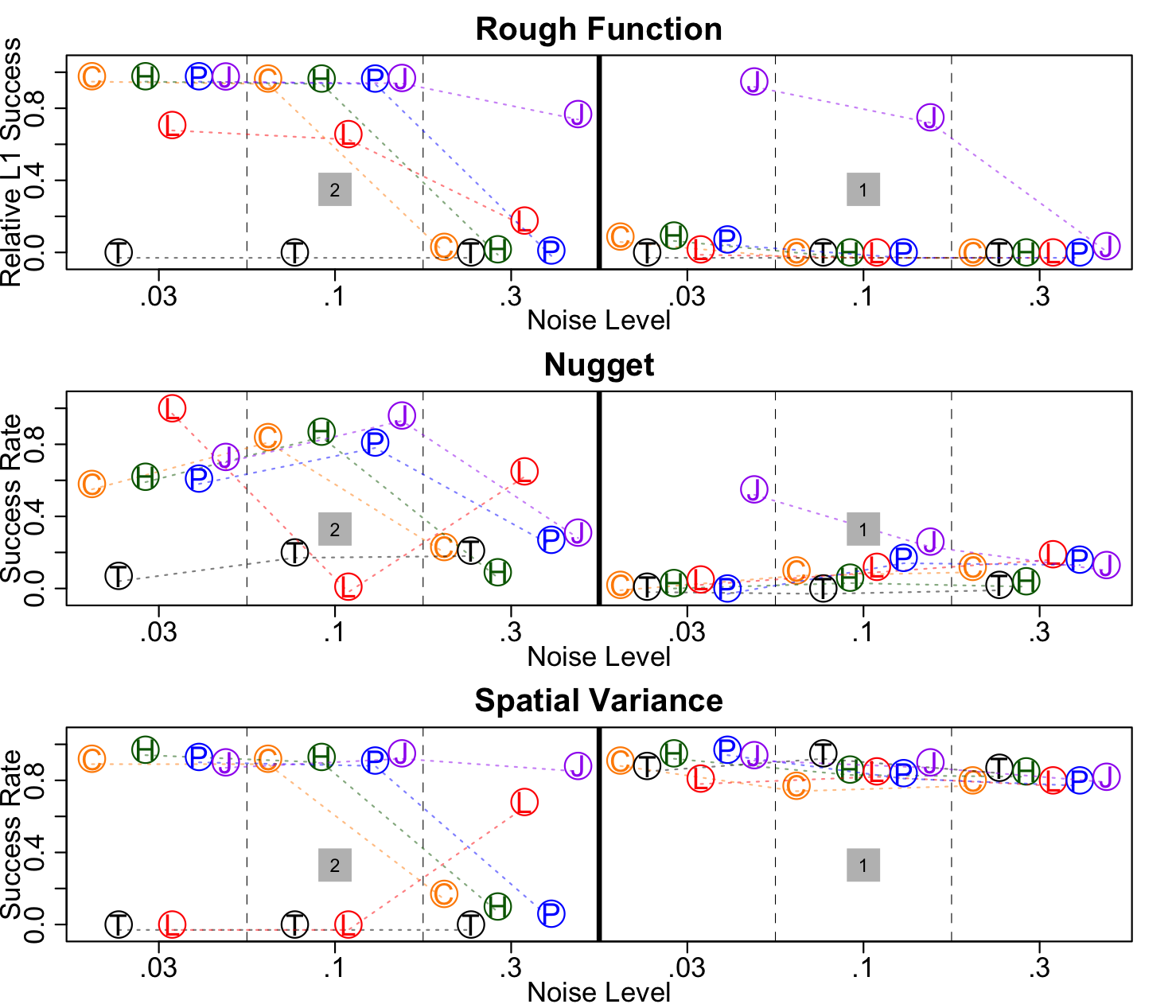}
    \caption{The success in recovering the true spatial parameters under a handful of the simulated situtations.  The numbers $2$ and $1$ represent the magnitude of the rough function.  T=TPS, L=LASSO, C=Cauchy, H=Horseshoe, P=Pareto scaled mixture, J=Normal Jeffrey's}
    \label{simstudresults}
\end{figure}

Figure \ref{synthetic.data} shows one sample of the synthetic data used in the simulation study.  The GP is generated using a TPS covariance function with a spatial variance, $\sigma^2=.5$. It is combined with a rough function containing plateaus of 0,1,2,3 units.
The results of this Monte Carlo study are summarized in Figure \ref{simstudresults}. We gauged the success rate of each model in identifying the true rough function by computing the relative error of the recovered function.
$$\text{Relative $L_1$ Success}=1-\frac{\|\mathbb{E}(\bgamma|z)-\bgamma\|_1}{\|\bgamma\|_1}$$
\noindent We also recorded the success rate of each scaling prior in identifying $\tau^2$ and $\sigma^2$ by recordin ghow often the true value of each was contained in a $95\%$ credible interval of the posterior. 

We found that the LASSO/Laplace prior is not an effective Bayesian tool on spatial data.  It fails to recover the rough function and, therefore, is unable to properly recover the linear effects, spatial effects, or the nugget.  The other scaled mixture priors do much better in identifying the true parameters.  Each was able to recover the true rough function under most conditions. The NJ prior is the most robust of the five. It was the only prior to properly estimate the true parameters in a number of more severe scenarios, including for all levels of noise when the magnitude of the rough function was $1$. We fit a TPS at each level to baseline our results.

\subsection{Distribution of the Scaling Parameters}

\begin{figure}
    \centering
    \includegraphics[scale=.16]{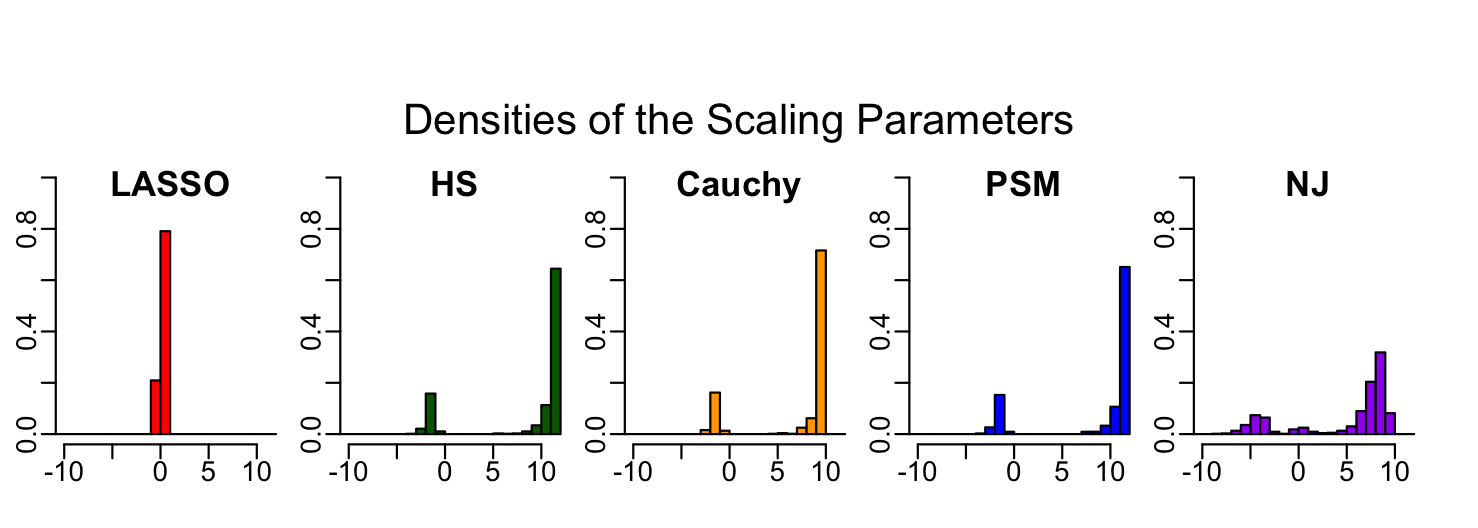}
    \caption{The distributions of $-log_{10}(\lambda^2_{\nu})$ under each scaling prior.}
        \label{penalties}
\end{figure}

The differences among the priors can be illustrated with histograms of posterior estimates of $\blambda^2$.  Recall that the inverse of the scaling parameter quantifies the level of fusion taking place between adjacent locations.  Figure \ref{penalties} contains histograms of $\frac{1}{\lambda^2_{\nu}}$, for each method in the scenario where the magnitude of the rough function is 2 and the noise level is $\tau=.1$.  Notice one mode for the  LASSO/Laplace prior, whereas the other scaled mixture priors separate the scaling parameters into two groups: one associated with fusion of adjacent points and one associated with step changes.

\section{Implementation}\label{Implementation}

\subsection{Orthogonalizing the BHM}\label{ortho}
The BHM in Equation \ref{int.BHM} suffers from confounding effects, so it is necessary to orthogonalize the fixed and spatial components \cite{hodges2010adding}. Without this transformation the features become unidentifiable and the MCMC fails to converge. \cite{hanks2015restricted} propose an effective orthogonalization technique by projecting each part of the model onto the basis functions of earlier components.  In our model, $\by$ lacks basis functions, so we add them using the covariance structure.  
\begin{align}
\bz=X\bbeta+\by+\bgamma+ \epsilon &\text{  where  }  \by\sim N(0,K\sigma^2) \\
&\text{   and   } \epsilon\sim N(0,I\tau^2) \nonumber 
\end{align} 
\noindent can be rewritten as the equivalent model
\begin{align}
\bz=X\bbeta+M\by+\bgamma+ \epsilon 
\end{align}    
\noindent where $\by\sim N(0,I\sigma^2)$ and $M$ is the Cholesky factor of $K$. M is a set of basis functions for $\by$ to orthogonalize $\bgamma$ against.\\
\\
Let $\bP_X=X(X^TX)^{-1}X^T$. Then
\begin{align}
    \bz&=X\bbeta+M \by+ \bgamma + \epsilon \nonumber \\
    &=X\bbeta+P_\bX(M \by+\bgamma) +(I-P_\bX)(M\by+ \bgamma)+ \epsilon \nonumber\\
     &= X\bbeta^*+ \Psi\by+ (I-P_\bX)\bgamma + \epsilon \nonumber
\end{align}
with  the reparameterization  $\Psi=(I-P_\bX)M$ and $\bbeta^* =\bbeta+(X^TX)^{-1}X^T(M\by+ \bgamma)$.
 Now, using the approach found in \cite{hofkes2024hybrid}, let 
 \[ P_{\Psi}=\Psi({\Psi}^T\Psi+I\delta)^{-1}{\Psi}^T\]
  where $\delta I$ functions as a small regularization term because $\Psi$ is rank deficient.  
\begin{align}
     \bz &= X\bbeta^*+ \Psi\by+ (I-P_\bX)\bgamma + \epsilon \nonumber \\
     &= X\bbeta^*+ \left( \Psi\by+ P_{\Psi}(I-P_\bX)\bgamma \right) + (I-P_{\Psi})(I-P_\bX)\bgamma +\epsilon \nonumber\\
    &= X\bbeta^*+ \Psi \by^* + H \bgamma +\epsilon
\end{align}
with the final reparametrizations 
 $\by^*=\by + J\bgamma$,  $J=({\Psi}^T\Psi+I\delta)^{-1} {\Psi}^T (I-P_\bX)$, and $H=(I-P_{\Psi})(I-P_X)$

\subsection{The Orthogonalized BHM}\label{BHMortho}

These modifications lead to an orthogonalized model for Bayesian hybrid smoothing of surfaces  
\begin{align}
    \bz|\bbeta^*,\by^*,\bgamma,\tau^2 &\sim N(X\bbeta^*+\Psi\by^*+H\bgamma,I\tau^2)  \\
    \by^*|\bgamma,\sigma^2 &\sim N(J\bgamma,I\sigma^2) \nonumber \\
    \bgamma|F(\blambda^2) &\sim N(0,F(\blambda^2)) \nonumber \\
    \pi(\blambda^2) & \text{ as a scaling distribution}\nonumber\\
     [\bbeta^*] &\propto 1 \nonumber\\
     \tau^2 & \sim \text{inv-gamma}(\alpha_{\tau^2},\beta_{\tau^2})  \nonumber \\
     \sigma^2 & \sim \text{inv-gamma}(\alpha_{\sigma^2},\beta_{\sigma^2})  \nonumber
 \end{align}

\subsection{Making $Q$ a Valid Precision}\label{validQ}
Remember, we do not write $F$ explicitly but, instead, build $Q$  as $D^T\Lambda^{-1}D+E$. On its own, $D^T\Lambda^{-1}D$ is not invertible, so additional information is needed, which we add in matrix $E$ via $\pi^*(\bgamma)$.
One option is to add an addition regularizing prior on $\bgamma$.
$$\pi^*(\bgamma) \sim N(\mathbf{0}, \frac{1}{\delta} I)$$ 
\noindent where $\delta$ is a small value, making $E=\delta I$. This prior adds $\delta$ to the diagonal of $D^T\Lambda^{-1}D$, making $Q$ invertible for large enough values of $\delta$. One downside of this approach, however, is that the ridge parameter, $\delta$, induces shrinkage in the individual $\bgamma$ values.   Another downside, it that the size of $\delta$ needed to make the precision computationally invertible, isn't immediately clear and will vary.  

An alternative to a ridge prior across all of $\bgamma$ is adding a very strong prior on exactly one $\bgamma$ location, something like $\bgamma_{center}^2\sim N(0,\text{1e-10})$ making $E$ a matrix of zeros with 1e10 on one diagonal entry.  This approach essentially defines one gamma location to be zero in the rough function.  While likely causing the MCMC to shift some of the overall mean to the rough function, this formulation anchors the rough functions, creates a valid precision, and results in an identifiable distribution for the sum of the overall mean and the rough function.  This is the approach we adopt.

Both of these approaches to making $Q$ invertible can also be introduced as a firm constraint instead of a prior, restricting $\bgamma$ to a subspace.  An additional constraint which would also stabilize $Q$ is $\mathbf{1}^T\gamma=0$.  This is the approach used in ANOVA.  These constraints will, however, effect the sparsity of $Q$, so we elect to use the prior information setup.

\subsection{Priors, Convergence, Adaptations}\label{setup}

\subsubsection{Priors and Convergence}
We used weakly informative, inverse gamma priors for $\tau^2$ and $\sigma^2$ with $\alpha_{\sigma^2}=\beta_{\sigma^2}=\alpha_{\tau^2}=\beta_{\tau^2}=.001$.  We found that Jeffrey's and uniform priors also work for $\sigma^2$. The model converges relatively quickly after a short burn-in of a few hundred and mixes well. We use a uniform prior on $\bbeta$ to maintain the correspondence between the GP and a cubic smoothing spline \cite{wahba1990spline}\cite{hofkes2024hybrid}.

\subsubsection{Adaptive Burn-in for the MCMC}\label{adaptive}
Poor initial conditions can lead to the MCMC getting stuck in suboptimal locations. The scale parameters, when left unchecked, have the tendency to fuse locations before $\bgamma$ can adequately adapt to the data.  This results in posterior samples of the rough function getting stuck around $\mathbf{0}$.  We avoid this problem by initially limiting the rate of fusion in the rough function.  

This is done by restricting the minimum value of draws from the conditional posterior of $\blambda^2$ and then slowly easing the restriction, ending it before the burn-in is completed.  This schedule allows the distribution of $\bgamma$ to be influenced by the data before fusing.  With this brake on $\blambda^2$, initial values are less consequential.

\subsubsection{Partial Updates}\label{partial}
The most expensive updates in the MCMC are the multivariate normal draws on GP, $\by$, and the NGP, $\bgamma$.  These parameters, however, are also the first to converge. Thus, the speed of overall convergence is greatly improved by skipping updates on $\by$ and $\bgamma$ on some passes through the MCMC, allowing the other parameters to catch up at a small computational cost.   We found that sampling $\by$ and $\bgamma$ every third pass through the MCMC was adequate to achieve convergence and an adequate effecting sample size on all parameters.

\section{The Gibbs Sampler}\label{Gibbs}
For completeness, we detail the Gibbs sampler for the orthogonalized BHM. 
\begin{align}
    \bbeta^*|\cdot &\sim N_n(A^{-1}b,A^{-1}) \\
    &\text{  where  } A=\frac{X^TX}{\tau^2} \text{  and  } b=\frac{X^T(\bz-\Psi\by^*-H\bgamma)}{\tau^2}\nonumber\\
    \by^*|\cdot &\sim N_n(A^{-1}b,A^{-1}) \text{  for  } A=\frac{I}{\sigma^2} + \frac{\Psi^T\Psi}{\tau^2}\\
    &\text{and  } b=\frac{\Psi^T(\bz-X\bbeta^*-H\bgamma)}{\tau^2} \nonumber \\
    \bgamma |\cdot & \sim N_n(A^{-1}b,A^{-1}) \text{  where  } A=\frac{H^TH}{\tau^2}+Q+\frac{J^TJ}{\sigma^2} \nonumber \\
    &\text{  and  } b=\frac{H^T(\bz-X\bbeta^*-\Psi\bg^*)}{\tau^2} +\frac{J^T\bg^*}{\sigma^2} \\
    \tau^2|\cdot & \sim \text{inv-gamma}\left(\frac{n}{2}+\alpha_{\tau^2}, r+\beta_{\tau^2}\right)\\
    \text{where } r&=\frac{\| \bz-X\bbeta^*-\Psi\by^*-H\bgamma\|^2}{2}  \nonumber  \\ 
    \sigma^2|\cdot  & \sim \text{gamma}\left(\frac{n}{2} + \alpha_{\sigma^2}, \frac{\|\by^*-J\bgamma\|^2}{2}+\beta_{\sigma^2}\right)
\end{align}

\noindent Additional parameter posteriors, corresponding to particular sparsity inducing priors, can be found in the subsections below.  Across all models we define $\lambda_{\nu}^2 =  \lambda_{\nu}^{*2} + \delta$ where $\delta$ is a lower limit on the values of $\blambda^2$.  This creates an upper limit on the scaling penalties, $\frac{1}{\blambda^2}$, and provides stability to $Q$. We use $\delta=$1e-12.

\subsection{Normal Jeffrey's - Jeffrey's Scaled Mixture}

The additional priors for this model are
\begin{align}
	 [ \lambda_{\nu}^{2*}] &\propto \frac{1}{ \lambda_{\nu}^{*2}}
\end{align}

\noindent The posteriors for this prior are
\begin{align}
	 \lambda_{\nu}^{*2}|\cdot &\sim \text{InverseGamma}\left(\frac{1}{2}, \frac{(\bgamma_{i(\nu)}-\bgamma_{j(\nu)})^2}{2}\right) 
\end{align}

\subsection{Horseshoe -  A Half-Cauchy Scaled Mixture of Half-Cauchy Scaled Mixtures}

The additional priors for this model are
\begin{align}
	 \lambda_{\nu}^2 &=  \lambda_{\nu}^{*2} + \delta \\
	 \lambda_{\nu}^*| t &\sim C^+(0,t)\\
	t &\sim C^+(0,\tau)
\end{align}
Using the identity in \cite{gelman2006prior}, we can rewrite each of the half-Cauchy distributions as two inverse gamma distributions
\begin{align}
	 \lambda_{\nu}^{*2}|v &\sim \text{InverseGamma}\left(\frac{1}{2},\frac{1}{v}\right)  \\
	v|t &\sim \text{InverseGamma} \left(\frac{1}{2},\frac{1}{t^2}\right)  \\
	t^2|a &\sim \text{InverseGamma}\left(\frac{1}{2},\frac{1}{a}\right)  \\
	a|\tau &\sim \text{InverseGamma} \left(\frac{1}{2},\frac{1}{\tau^2}\right)  
\end{align}

\noindent The posteriors for this prior are
\begin{align}
	 \lambda_{\nu}^{*2}|\cdot &\sim \text{InverseGamma}\left(1, \frac{(\bgamma_{i(\nu)}-\bgamma_{j(\nu)})^2}{2} + \frac{1}{v_{\nu}} \right) \\
	v_{\nu}|\cdot &\sim \text{InverseGamma}\left(1,\frac{1}{\lambda^{*2}_{\nu}} + \frac{1}{t^2} \right) \\
	t^2 &\sim \text{InverseGamma}\left(m+1, \sum_{\nu}\frac{1}{v} + \frac{1}{a^2}\right)\\
	a &\sim \text{InverseGamma}\left(\frac{1}{t^2}+\frac{1}{\tau^2}\right) \\
	&\text{Additionally, the conditional posterior on $\tau^2$ changes to } \nonumber\\
	\tau^2|\cdot &\sim \text{inv-gamma}\left(\frac{n+1}{2}+\alpha_{\tau^2}, r+\beta_{\tau^2}+\frac{1}{a}\right)\\
    	\text{where } r&=\frac{\|\bz-X\bbeta^*-\Psi\by^*-H\bgamma\|^2}{2} \nonumber
\end{align}

\subsection{LASSO - Exponential Scaled Mixture}

The additional priors for this model are
\begin{align}
	 \lambda_{\nu}^2 &=  \lambda_{\nu}^{*2} + \delta\\
	 \lambda_{\nu}^{*2} | b^2 &\sim exp(b^2/2)\\
	 [b^2] &\propto \frac{1}{b^2}
\end{align}

\noindent The posteriors for this prior are
\begin{align}
	\frac{1}{ \lambda_{\nu}^{*2}}|\cdot \sim \text{InverseGaussian}\left(\sqrt{\frac{b^2}{(\bgamma_{i(\nu)}-\bgamma_{j(\nu)})^2}},b^2\right) \\
	b^2 \sim \text{Gamma}\left(m,\sum_{\nu} \lambda^2_{\nu}/2\right)
\end{align}

\noindent It is important to note that the posterior of this exponential scaled mixture is not unimodal.  Although the $L_1$ LASSO penalty is convex, this scaled mixture introduces a latent variable, $\blambda^2$, and the derivative of the log posterior with respect to $\blambda^2$ has a quadratic form with multiple zeros.  This is not necessarily problematic for Bayesian methods, but it needs to be paid attention to when assessing convergence.

\subsection{Cauchy - Inverse Gamma Scaled Mixture}

The additional priors for this model are
\begin{align}
	 \lambda_{\nu}^2 &=  \lambda_{\nu}^{*2} + \delta\\  
	 \lambda_{\nu}^*| b &\sim \text{InverseGamma}\left(\frac{1}{2},\frac{1}{2b^2}\right)\\
	[b^2] &\propto \frac{1}{b^2}
\end{align}

\noindent The posteriors for this prior are
\begin{align}
	 \lambda_{\nu}^{*2}|\cdot &\sim \text{InverseGamma}\left(1, \frac{(\bgamma_{i(\nu)}-\bgamma_{j(\nu)})^2}{2} + \frac{1}{2b^2} \right) \\
	b^2 &\sim \text{InverseGamma}\left(\frac{m}{2}, \sum_{\nu}\frac{1}{2 \lambda_{\nu}^{*2}}\right)
\end{align}

\subsection{Pareto Scaled Mixture}

The additional priors for this model are
\begin{align}
	 \lambda_{\nu}^2 &=  \lambda_{\nu}^{*2} + \delta \\
	 \lambda_{\nu}^*| \alpha, \lambda^2_{min} &\sim Pareto(\alpha,\lambda^2_{min})\\
	 [\alpha] & \propto \frac{1}{\alpha}\\
	 [\lambda^2_{min}] &\propto \frac{1}{\lambda^2_{min}}
\end{align}
\noindent The posteriors for this prior are
\begin{align}
	 \lambda_{\nu}^{*2}|\cdot &\sim \text{InverseGamma}\left(1+\alpha, \frac{(\bgamma_{i(\nu)}-\bgamma_{j(\nu)})^2}{2}\right) \\
	\alpha &\sim exp\left(\sum_{\nu} log( \lambda_{\nu}^{*2}) - m log(\lambda_{min})\right)\\
	\lambda^2_{min} &= exp\left(\frac{log(U)}{m\alpha}+log(min(\blambda^{*2}))\right)\\
	&\text{where } U\sim Uniform(0,1) \nonumber
\end{align}

\noindent Note that $\lambda^2_{min}$ is being sampled directly. See Appendix \ref{pareto} for the derivation of the Pareto parameter posteriors.

\section{Climate Sensitivity Data}\label{CSD}
The Community Earth System Model (CESM) was developed by the National Center for Atmospheric Research (NCAR) in collaboration with other institutions and university researchers.  It is an open source model, sponsored by the National Science Foundation (NSF) and freely available to the research community.  As an example, we focus on the derived surface field of local climate sensitivity.  This is the local change in temperature based on a one degree change in global temperature.  CESM-LE is comprised of 30 simulations of internal climate variability created to aid in the assessment of climate change. \cite{kay2015community}. \\
\indent A number of papers propose GP models to emulate the LENS output fields.  \cite{wiens2020modeling} considers the additive model for climate sensitivity as
\begin{align}
z(\bs) = \mu(\bs)+y(\bs)+\epsilon(\bs)
\end{align}
\noindent where $\epsilon$ is mean zero white noise, $\by$ is a mean zero Gaussian process, $\mu$ is the expected set of values for any realization, and $\bs$ is the set of spatial locations.  \cite{wiens2020modeling} estimated $\mu$ as a point-wise mean of 30 ensemble members. \\
\indent Here, we estimate $\bmu$ in the model using hybrid smoothing.  We compare this result and its associated uncertainty with a Bayesian model on all 30 realizations and a thin plate spline on the pointwise means of the data.  We do this by adapting the top line of the hierarchical model in Section \ref{BHM} to use all $30$ members of the ensemble.
\begin{align}
    \bz_i|\bbeta^*,\by^*,\bgamma,\tau^2 &\sim N(X\bbeta^*+\Psi\by^*+H\bgamma,I\tau^2)
 \end{align}
\noindent This change produces new posterior distributions on a number of parameters.  The adjustments in the Gibbs Sampler are as follows. Let $m$ be the number of realizations included in the data, and $\bz_{(i)}$ be the $i^{th}$ realization of the data.
\begin{align}
    \bbeta^*|\cdot &\sim N_n(A^{-1}b,A^{-1}) \text{  where  } A=\frac{mX^TX}{\tau^2} \\
    & \text{  and  } b=\sum_{i=1}^m \frac{X^T(\bz_{(i)}-\Psi\by^*-H\bgamma)}{\tau^2} \nonumber\\
    \by^*|\cdot &\sim N_n(A^{-1}b,A^{-1}) \text{  for  } A=\frac{I}{\sigma^2} + \frac{m\Psi^T\Psi}{\tau^2}\\
    &\text{and  } b=\sum_{i=1}^m\frac{\Psi^T(\bz_{(i)}-X\bbeta^*-H\bgamma)}{\tau^2} \nonumber \\
    \bgamma |\cdot & \sim N_n(A^{-1}b,A^{-1}) \text{  where  } A=\frac{H^TH}{\tau^2}+Q+\frac{mJ^TJ}{\sigma^2} \nonumber \\
    &\text{  and  } b=\sum_{i=1}^m\frac{H^T(\bz_{(i)}-X\bbeta^*-\Psi\bg^*)}{\tau^2} +\frac{J^T\bg^*}{\sigma^2} \\
    \tau^2|\cdot & \sim \text{inv-gamma}\left(\frac{mn}{2}+\alpha_{\tau^2}, r+\beta_{\tau^2}\right)\\
    \text{where } &=\sum_{i=1}^m\frac{\|\bz_{(i)}-X\bbeta^*-\Psi\by^*-H\bgamma\|^2}{2}  \nonumber
\end{align}

\subsection{Results on Climate Sensitivity Data}

We implemented the Bayesian hybrid smoother for each of the five different priors listed in Table \ref{scaled} and also fit a Bayesian TPS model to the ensemble. Our goal was for the hybrid smoother to capture the change in climate sensitivity between land and water, leading to more accurate modeling of the mean function with fewer effective degrees of freedom (EDF).  We included elevation as a covariate in the models to avoid identifying changes solely in this variable. Given the level of noise in climate sensitivity data, we were uncertain whether any of the priors would adequately capture this shift.

\begin{figure}
    \centering
    \includegraphics[scale=.16]{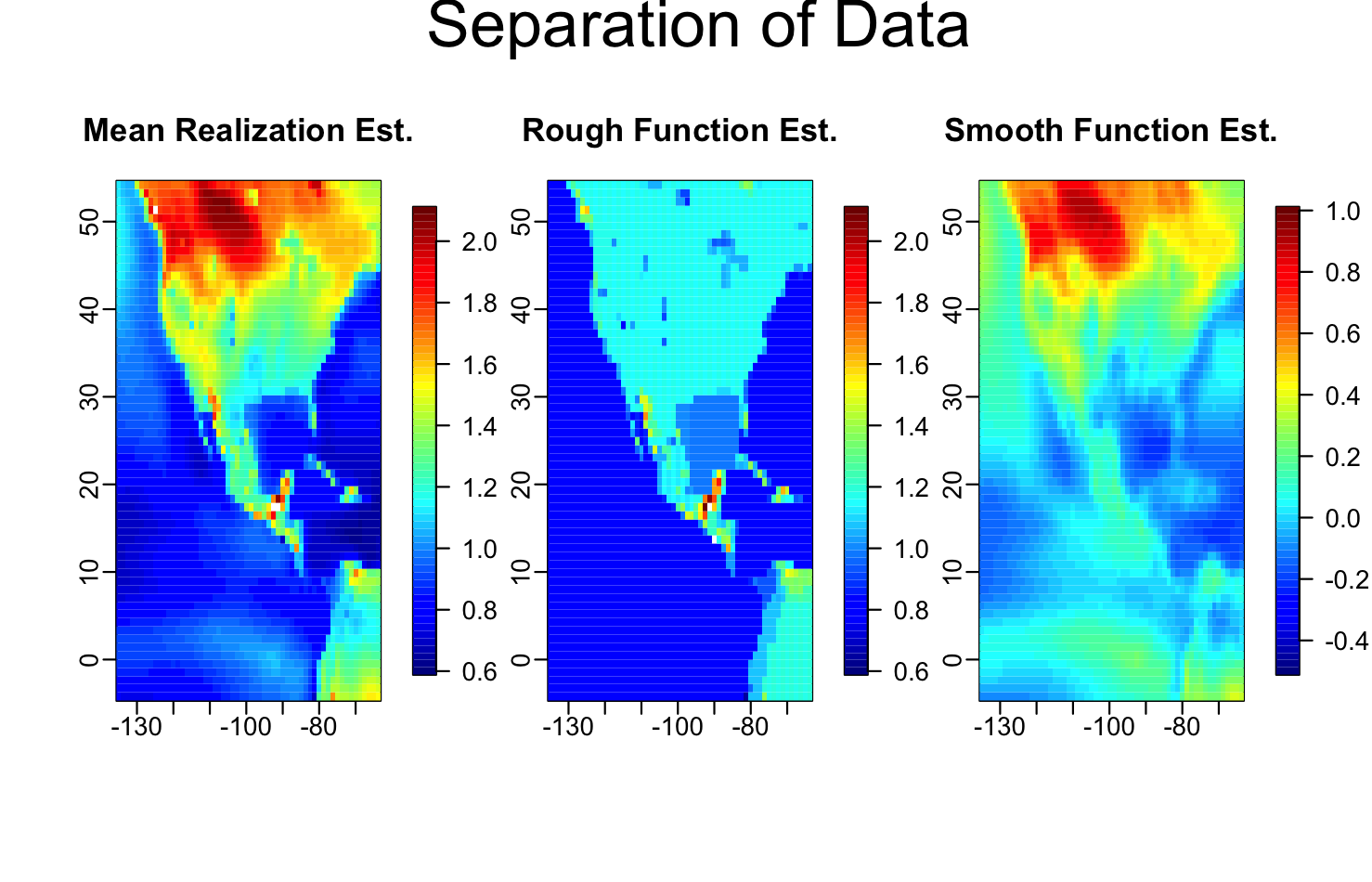}
    \vspace{-2.1em}
    \caption{The results of using hybrid smoothing with a NJ prior on the derivative of the rough function.}
    \label{results.climate}
\end{figure}

The LASSO failed to identify any rough function at all, while the Horseshoe, Cauchy, and Pareto-scaled mixtures identified some rough features but only extreme ones in a few mountainous regions near the coast. In contrast, the NJ prior successfully identified an interpretable rough signal, delineating land and ocean areas. It also distinguished the Gulf of Mexico from the Pacific and Atlantic Oceans and identified the Great Lakes. The distribution of scaling parameters associated with the NJ prior appears crucial in inducing fusion in some areas while avoiding it in others. Figure \ref{results.climate} shows the results of the NJ hybrid BHM.

\begin{figure}
    \includegraphics[scale=.16]{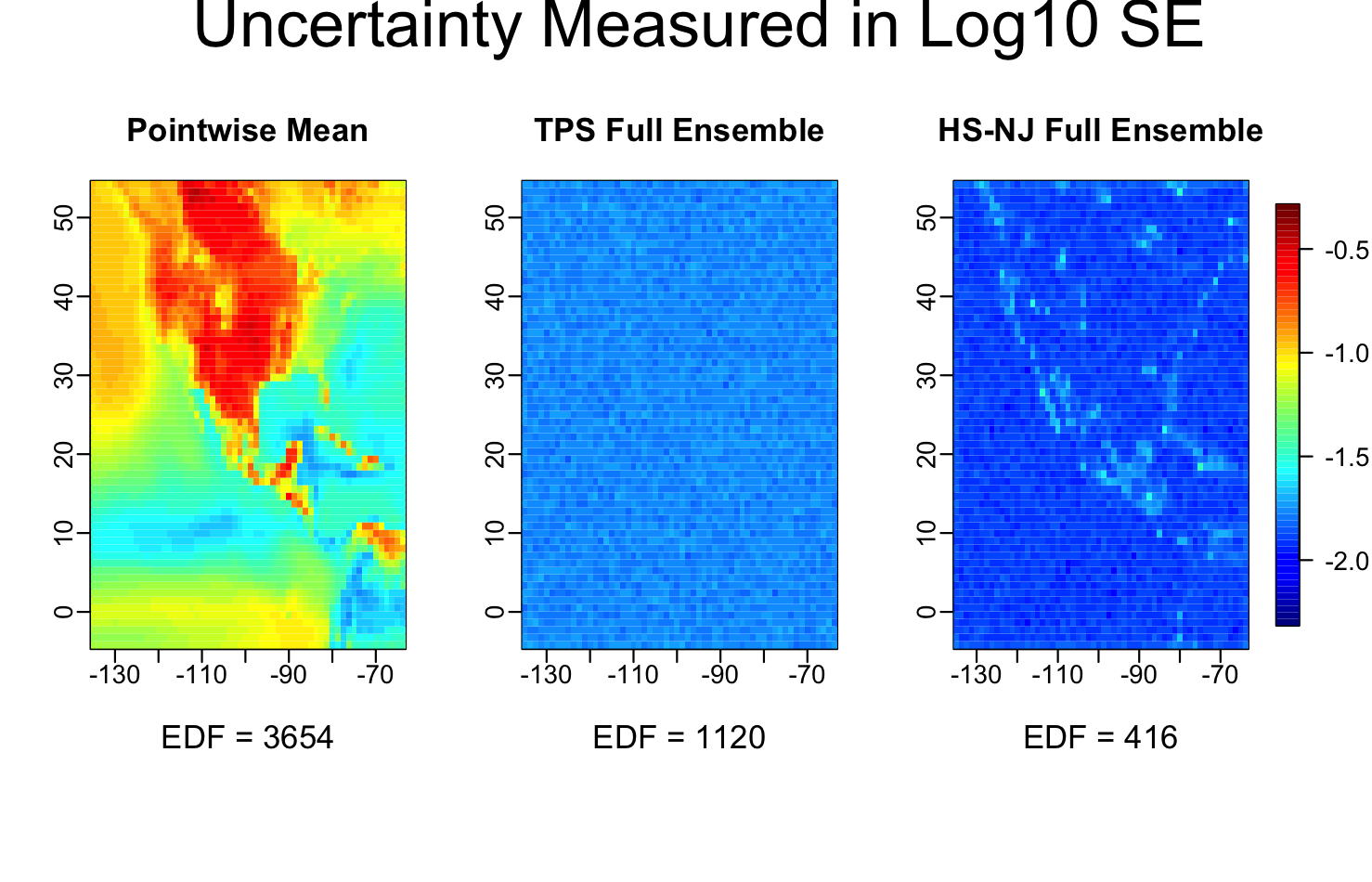}
    \centering
    \vspace{-2.1em}
    \caption{The log uncertainty associated with each method. Below each plot in the EDF associated with the method.}
    \label{uncertainty}
    \vspace{-1em}
\end{figure}

In addition to $\mu(\bs)$, another quantity of interest is the uncertainty in $\mu(\bs)$ across realizations.  We compare the uncertainty of the hybrid Bayesian approach using all thirty ensemble members with the uncertainties associated with the traditional spatial BHM, a pointwise estimate of the mean, and a thin plate spline, fit on the average values at each location.  Figure \ref{uncertainty} displays the results.  The size of the uncertainty between methods varies greatly.  We see that both of the Bayesian model estimates with the full ensemble have significantly less uncertainty than the other two approaches.  It should be noted that smaller uncertainty is not necessarily a sign of more accurately discovering the true function, however, and these values should be judged in relation to the EDF associated with each model.  Models with large EDF risk over-parametrization and poor fit to the true function.  The Bayesian models with the full ensemble exhibit less uncertainty in far fewer EDF than the previous approaches.  The hybrid approach using a NJ prior on the rough function dramatically decreases the EDF even further.

\section{Extensions to the model}\label{Extensions}

The factorization of $Q-E$, leads to an easy extension of the model to other linear features.  Instead of extending the mean to a piecewise step function, it could extend the linear components of the model to a piecewise linear function by swicthing $D$  from the first discrete differencing matrix to the second or third discrete differencing matrix. 
\[ \sum_{\nu} \frac{1}{\lambda_i^2} \left( \gamma_{i(\nu)}-2 \gamma_{j(\nu)}+ \gamma_{k(\nu)}\right)^2 = \bgamma^T ( D^T \Lambda^{-1} D) \bgamma\]
\[ \sum_{\nu} \frac{1}{\lambda_i^2} \left( \gamma_{i(\nu)}-3 \gamma_{j(\nu)} + 3\gamma_{k(\nu)}-\gamma_{l(\nu)}\right)^2 = \bgamma^T ( D^T \Lambda^{-1} D) \bgamma  \]

where $k(\nu)$ and $l(\nu)$ output the field location corresponding to the additional element used to compute the second and third discrete differences respectively.  These increases in differencing order increase the rank deficiency of $D^T \Lambda^{-1} D$ and therefore do require additional prior information or constrains. This structure merits further study, not only due to its value in change point detection, but also the sparse nature of the precision matrix, making it well-suited for large spatial data, even for larger order differences.

\section{Conclusion}\label{Conclusion}

In this paper, we addressed the challenge of modeling spatial data with both smooth and rough processes, where traditional Kriging and non-parametric spatial models often fall short, particularly when covariates are unavailable. We introduced a hybrid smoothing approach, combining a non-Gaussian process (NGP) modeled as a scaled mixture of Gaussian Processes (GP) to induce sparsity on the rough surface, with a standard GP to capture smooth features.  This approach reduces the EDF and provides improved uncertainty bounds.

Our simulation study demonstrated the robustness of hybrid smoothing, showing that it can accurately recover spatial parameters in the presence of step changes. By testing the five scaled mixtures over a range of step changes and noise levels, we explored the limits of the method and found various levels of robustness associated with each mixture.  We illustrated how these levels of robustness are connected to the ways in the which the mixtures shrink the differences in adjacent locations.

The Normal Jeffrey's mixture was particularly well suited for identifying and simulating rough spatial processes in the CESM-LE and Monte Carlo study. Its lack of hyper-parameters is noteworthy, reducing subjectivity in the inference process.

The application to the CSEM-LE climate sensitivity data highlighted the model's ability to improve both the estimate of the mean function and the uncertainty associated with it.  By implementing advanced techniques such as adaptive burn-ins and orthogonalization, we further enhanced the computational efficiency of the BHM.

Despite the improvements achieved, the computational complexity remains a consideration for large datasets.  The sparse nature of the precision associated with the rough spatial process appears to make the NGP a straightforward addition to large spatial models that exploit sparsity.  Sparse methods also provide a framework for switching the spatial dependency in $\bgamma$ from the data to a set of basis function coefficients.  This would allow for irregularly spaced data. 

Overall, hybrid smoothing represents a significant advancement in spatial modeling, providing a flexible and effective approach for modeling spatial processes with a complex mean structure.

\bibliographystyle{IEEEtran}
\bibliography{Bibliography-surface.bib}

\section*{APPENDIX}

\section{The Posterior of Pareto Parameters}\label{pareto}

The full posterior of the additional parameters in this model, where $\lambda^2_{\nu} > \lambda^2_{min}$,  is

\begin{align}
[\blambda^2,\alpha,\lambda^2_{min}] \propto  \prod_{\nu} \left(\frac{\alpha(\lambda^2_{min})^{\alpha}} {(\lambda^2_{\nu})^{(\alpha+1)}} \right) \left(\frac{1}{\alpha}\right)\left(\frac{1}{\lambda^2_{min}}\right) 
\end{align}

\noindent The posterior on $\alpha$: 
\begin{align}
[\alpha|\cdot] & \propto \prod_{\nu} \left(\frac{\lambda^2_{min}}{\lambda^2_{\nu}}\right)^\alpha \\
& \propto \exp\left(-\alpha \left(\sum_{\nu} log(\lambda^2_{\nu}) - log(\lambda^2_{min})\right)\right)\\
\alpha|\cdot &\sim \text{Exp}\left(\sum_{\nu} log(\lambda^2_{\nu}) - log(\lambda^2_{min})\right)
\end{align}
\\
\noindent The posterior on $\lambda^2_{min}$: 
\begin{align}
[\lambda^2_{min}|\cdot] \propto (\lambda^2_{min})^{m(\alpha-1)}
\end{align}
Computing the normalizing factor, we get 
\begin{align}
[\lambda^2_{min}|\cdot] = \frac{m\alpha}{min(\blambda^2)^{m\alpha}} (\lambda^2_{min})^{(m\alpha-1)}
\end{align}
By inverse transform sampling, with $U\sim \text{Uniform}(0,1)$,
\begin{align}
\lambda^2_{min}|\cdot=\sqrt[m\alpha]{U min(\blambda^2)^{m\alpha}}
\end{align}
For stability, we rewrite the above sampling formula as
\begin{align}
	\lambda^2_{min}|\cdot &= exp\left(\frac{log(U)}{m\alpha}+log(min(\blambda^{2}))\right)
\end{align}

\section{Sampling from Jeffrey's Portion of Normal Jeffrey's Prior} \label{jeffreys}

We can sample from Jeffrey's prior on $blambda^2$ in the following manner.  Set upper and lower limits on $\blambda^2$, such as 1e100 and 1e-100 respectively. The CDF of $\blambda^2_i$ is then
\begin{align}
	\mathbf{F}(\blambda^2_i)\approx  \frac{ \displaystyle \int_{\text{lower}}^{\blambda^2_i} \frac{1}{x} dx   }{  \displaystyle \int_{\text{lower}}^{upper} \frac{1}{x}dx} = \frac{log(\blambda^2_i)-log(\text{lower})}{log(\text{upper})-log(\text{lower})}
\end{align}
\noindent This distribution can be sampled from using inverse transform sampling. Let $U \sim \text{U}(0,1)$.
\begin{align}
\blambda_i^2=e^{(U) log(\text{upper})+(1-U)log(\text{lower})}
\end{align}

\noindent This distribution is uniform on a log scale.

\end{document}

%% file: ourDefinitionsLite.tex
\DeclareMathOperator*{\argmin}{\arg\!\min}

\newcommand{\bg}{\mathbf{g}}

\newcommand{\bs}{\mathbf{s}}

\newcommand{\by}{\mathbf{y}}
\newcommand{\bz}{\mathbf{z}}

\newcommand{\bP}{\mathbf{P}}

\newcommand{\bX}{\mathbf{X}}

\newcommand{\bbeta}{\boldsymbol{\beta}}
\newcommand{\bepsilon}{\boldsymbol{\epsilon}}
\newcommand{\bgamma}{\boldsymbol{\gamma}}

\newcommand{\blambda}{\boldsymbol{\lambda}}

\newcommand{\bmu}{\boldsymbol{\mu}}